\documentclass[12pt,dvips,a4paper,leqno]{article}

\usepackage{german}
\usepackage[latin1]{inputenc}
\usepackage[T1]{fontenc}
\usepackage{parskip}

\usepackage{amsmath,amsfonts,amssymb,amsxtra}
\usepackage{latexsym}
\usepackage{theorem}
\usepackage{graphicx,psfrag}
\usepackage{verbatim}

{\theoremstyle{plain}                       
\setlength{\theorempreskipamount}{1cm}
\setlength{\theorempostskipamount}{1cm}
\theorembodyfont{\itshape}
}

{\theoremstyle{plain}                       
\theorembodyfont{\itshape}
}

{\theoremstyle{plain}                       
\theorembodyfont{\itshape}
}

{\theoremstyle{plain}                       
\theorembodyfont{\itfamily}
}

{\theoremstyle{plain}                       
\theorembodyfont{\rmfamily}
}

{\theoremstyle{plain}                       
\theorembodyfont{\rmfamily}
}

{\theoremstyle{plain}                       
\theorembodyfont{\rmfamily}
}

\begin{document}    

 \begin{center}
  \Large{QED - classically. An intuitive Approach}\\
  \vspace{1cm}
  \large{Tibor Dudas}
 \end{center}

 \section*{Brief introduction}
 This paper deals with QED-particles and the interaction between them on a classical level.
 The Maxwell-equations are used mainly. (Proofs are not used in a mathematical but intuitive sense.)
 In the first step the main statements are presented. The corresponding proofs are given in the 
 second and final step.

 \section*{Statements}
 \begin{itemize}
  \item[I.]   An electron/positron is a sink/source of electromagnetic scalar quanta. 
  \item[II.]  These electromagnetic scalar field quanta should clearly be identified with photons.
  \item[III.] Charge-charge interaction is easily conceivable.
  \item[IV.]  The magnetic vector field constitutes a flow of photons.
  \item[V.]   It becomes clear, why a charged particle is deflected perpendicular to the magnetic field lines. 
 \end{itemize}

 \section*{Proofs}
  
 \noindent\textbf{I.} \
 \textsc{Lemma.} \
 It is sufficient to consider the electromagnetic scalar field only without the vector field.

 \noindent\textsc{Proof.} \
 According to the Lorentz-condition
 \begin{equation*}
  \partial_{\mu} A^{\mu} = 0
  \quad \Leftrightarrow \quad
  \partial_0 A^0 + \vec{\nabla}\vec{A} = 0  
 \end{equation*}
 It is 
 \begin{equation*}
 A^{\mu} =
 \left(
  \begin{array}{c}
   \Phi \\
   \vec{A}
  \end{array} 
 \right)
 \end{equation*}
 and
 \begin{equation*}
  x^{\mu} = 
  \left(
   \begin{array}{c}
    t \\
    \vec{x}
   \end{array} 
  \right)
 \end{equation*}
 with 
 \begin{eqnarray}
  \vec{B} & = & \vec{\nabla}\times\vec{A} \quad \mbox{and} \\
  \vec{E} & = & -\vec{\nabla}\phi-\frac{1}{c}\partial_t\vec{A}. 
 \end{eqnarray}
 
 Integrate the Lorentz-condition over a finite volume and use Gauss' law in order to obtain
 \begin{eqnarray*}
  O & = & \int\limits_{V} \partial_t \phi d^{\, 3}x +
          \int\limits_{V} \vec{\nabla} \vec{A} d^{\, 3}x 
      =   \int\limits_{V} \partial_t \phi d^{\, 3}x+
          \int\limits_{\partial V} \vec{A} d\vec{n} \\
    & \Leftrightarrow &
          \int\limits_{V} \partial_t \phi d^{\, 3}x =
          -\int\limits_{\partial V} \vec{A} d\vec{n}           
 \end{eqnarray*}
 \begin{figure}[h] 
  \begin{center}
   \includegraphics[width=0.4\columnwidth]{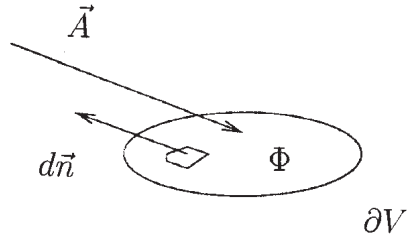} 
  \end{center}
  \begin{center}
   \parbox{8cm}{\caption{If $\Phi$ changes with time, it is due to a flow of $\Phi$ into
           or out of the volume.}}  
  \end{center} 
 \end{figure}

 Now, if $\phi$ is increasing with time, there is a net inward flow or a net nonvanishing component of 
 $\vec{A}$ pointing inward the volume.\hfill$\Box$ 

 Notice: \  $\partial_t \rho + \vec{\nabla} \vec{\jmath} = 0$ for charge density $\rho$ and 
 $\vec{\jmath} = \rho\vec{v}$ being the charge flux.

 Hence, let us make the following asumptions:
 (1) $\vec{A}$ is a flow of the electromagnetic scalar field $\phi$.
 (2) $\phi$ is quantized.
 
 \noindent\textsc{Proof (I. Statement).} \
 For a point charge $\phi\sim\frac{1}{r}$.

 The scalar potential should weaken, if the seperation $ds$ between its quanta increases.
 Therefore, we expect
 \begin{equation*}
  \phi \sim \frac{1}{ds} = \frac{1}{r},  
 \end{equation*}
 which is a confirmation of our image.
 \begin{figure}[h] 
  \begin{center}
   \includegraphics[width=0.4\columnwidth]{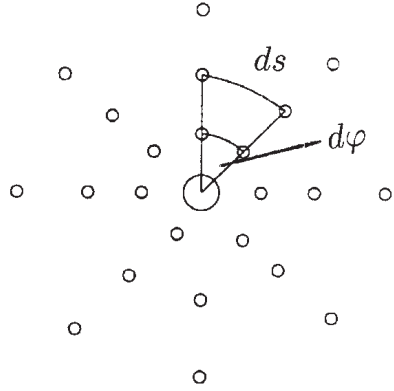} 
  \end{center}
  \begin{center}
   \parbox{8cm}{\caption{The photons \ flow/emerge \ into/from \ the electron/positron. 
            $\Phi\sim\frac{1}{ds}=\frac{1}{r}$. The scalar potential 
            should weaken, if the seperation $ds$ between its quanta 
            increases.}}
  \end{center} 
 \end{figure}

 According to $\vec{E}=-\vec{\nabla}\phi$ for an electron the concentration increases for bigger 
 distances from the core. Hence, an electron  is a sink. By reverse argumentation, a positron
 is a source of electromagnetic scalar field quanta.\hfill$\Box$
The scalar potential should weaken, if the seperation $ds$ between its quanta increases.

 \noindent\textbf{II. } \
 Trivial, as nescessarily, these quanta represent the photons, as they create the 
 electric field via (2).
 \hfill$\Box$

 \noindent\textbf{III. } \
 Charge-Charge interaction.

 \noindent 
 (1) Electron-Photon. The moving electron is dragged towards to source of electromagnetic scalar field and vice versa.

 \noindent
 \textsc{Proof.} \ This is a fact, just the image is different. 

 \noindent
 (2) Electron-Electron. The agglomeration of photons between the two sources pushes them apart
 from each other, the same way like two firemen sitting on movable chairs would seperate by
 directing their running waterguns towards each other. 

 \noindent
 (3) As the concentration of electromagnetic scalar field quanta between the two sinks is lower than aside from them,
 they are dragged apart towards the higher concentration, therefore seperate.\hfill$\Box$

 \noindent\textbf{IV. } \
 Magnetism/Lorentz-force.

 $\vec{B}=\vec{\nabla}\times\vec{A}$ and $\vec{A}$ is flow of eletromagnetic scalar field quanta (see Lemma in I).
 Therefore, we have a nonvanishing whirl of moving photons, constituting the magnetic field.
 \hfill$\Box$

 \noindent\textbf{V. } \
 Lorentz-force
 \begin{center}
  \begin{figure}[h] 
   \begin{center}
    \includegraphics[width=7cm]{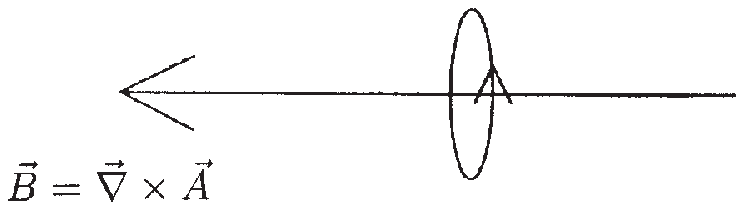} 
   \end{center}
   \begin{center}
    \parbox{8cm}{\caption{A whirl of electromagnetic scalar field $\Phi$ quanta constitute 
             the magnetic field. 
             $-\int\limits_{\partial V} \vec{A} d\vec{n} =
             \int\limits_{V} \partial_t \phi d^{\, 3}x $.}}
   \end{center}
  \end{figure}
 \end{center}
 \begin{equation*} 
  \vec{B}=\vec{\nabla}\times\vec{A}, \qquad \vec{F}=q\vec{v}\times\vec{B}.
 \end{equation*}
 $\mbox{rot}\vec{A}$ is a whirl of moving photons.
 An electron is deflected downwards coming from the front, as it runs towards, where the most
 photons are coming from.
 A positron is deflected upwards, as photons are coming slightly from below.

 This is what the Maxwell equations say together with the Lorentz condition taken physically
 relevant. 

 Magnetism can therefore be reduced to photon-charge interaction in the same way like charge-charge 
 interaction takes place, classically. \hfill$\Box$

\end{document}